\documentclass[structabstract]{aa}  
\usepackage{graphicx}
\usepackage{natbib}
\usepackage{txfonts}
\begin{document}
   \title{The complex structure of the disk around  HD~100546
\thanks{Based on observations collected at the VLTI
   (ESO Paranal, Chile) with program 076.C-0851(B)}} 
   \subtitle{The inner few astronomical units.}
   \author{M.~Benisty\inst{1} 
          \and 
          E.~Tatulli\inst{2}
          \and F.~M\'enard\inst{2}
          \and M.R.~Swain\inst{3}} 
        \offprints{benisty@arcetri.astro.it}
        \institute{INAF-Osservatorio  Astrofisico  di  Arcetri,  Largo
          E.~Fermi 5, 50125 Firenze, Italy
          \and
          Laboratoire d'Astrophysique de  Grenoble, CNRS-UJF UMR 5571,
          414 rue de la Piscine, 38400 St Martin d'H\`eres, France
          \and
          Jet   Propulsion   Laboratory,   California   Institute   of
          Technology, 4800 Oak Grove Drive, Pasadena, CA 91109, USA\\}

        \date{Received 03/11/2009; Accepted 07/01/2010}

                                % \abstract{}{}{}{}{} 
                                % 5 {} token are mandatory
   \abstract
   {Disclosing the structure of disks surrounding Herbig~AeBe stars is
   important to  expand our understanding  of the formation  and early
   evolution  of stars and  planets. The  first astronomical  units of
   these disks in particular, because they are hot, dense, and subject
   to intense  radiation field, hold  critical clues to  accretion and
   ejection  processes, as  well  as planet  formation in  environment
   different than what prevailed around our own early Sun.}  
   {We aim at revealing the sub-AU disk structure around the 10~Myr old
     Herbig~Be star  HD~100546 and at investigating the origin of its
     near and mid-infrared excess.}  
   {We used new AMBER/VLTI observations to resolve the K-band emission
     and to constrain the location  and composition of the hot dust in
     the  innermost circumstellar  disk. Combining  AMBER observations
     with photometric  and MIDI/VLTI measurements from  the litterature, we
     revisit the disk geometry using  a passive disk model based on 3D
     Monte-Carlo   radiative  transfer  (including   full  anisotropic
     scattering). }%  The calculations  reproduce well  the visibilities
                                %     and the entire spectral energy distribution.}  
   {We propose a  model that includes a tenuous inner disk
     made of micron-sized dust grains,  a gap, and a massive optically
     thick    outer   disk,    that   successfully    reproduces   the
     interferometric data and the SED.  We locate  the   bulk  of
     the  K-band   emission  at $\sim$0.26~AU. Assuming that this
     emission originates from  silicate dust grains at their sublimation
     temperature of $1500$K, we show that micron-sized grains are
     required to enable the dust to survive at such a close distance
     from the star. As a consequence, in our
    best  model, more  than 40\%  of the  K-band flux  is  related to
    scattering, showing that the direct thermal emission of  hot dust
    is not always sufficient  to explain the near-infrared excess.  In
     the  massive  outer  disk,  large  grains in  the  mid-plane  are
     responsible for the mm emission while a surface layer of 
    small  grains allows  the  mid  and far  infrared  excesses to  be
     reproduced.  Such  vertical  structure  may be  an  evidence  for
     sedimentation.  The interferometric observations
     are consistent with a disk model that includes a gap until $\sim 13$~AU from
     the  star  and  a  total  dust mass  of  $\sim$0.008  lunar  mass
     ($\sim 6.10^{23}$ g) inside
     it.  These values together with the derived scale
     height ($\sim 2.5$~AU) and temperature ($\sim$220~K) at the inner
     edge of the outer disk (r$=$13~AU), are consistent with recent CO~observations.}

\keywords{Stars  (individual):  HD  100546 -  Techniques:  Infrared
   interferometry} 
 \maketitle

\section{Introduction}
Extrasolar planets have been discovered at a regular pace since 1995.
Late in 2008, five of them were imaged directly, two of them clearly
being associated with the disk they formed in \citep{kalas, lagrange}. Yet, the exact process
by which they form, and the timing of their formation within the
disks remain unclear. There has been several interesting advances in
the last years: direct imaging of debris disks, the
discovery of fine structure in young disks likely due to
gravitational interactions with bodies  within or outside the disk,
and finer sampling (spectroscopy) of the mid-infrared spectral energy
distribution (SED).  It is now clear that several processes are
simultaneously acting on the disks at a given time. In the same time interval, it is 
expected that planets are actively forming.

In     a     series     of    multi-scale     panchromatic     studies
\citep{doucet_1,pinte_5,grady_3}, we have initiated the study of 
several well documented disks  surrounding pre-main-sequence stars. They will provide a 
testbed to study the conditions in which planets form.  
One of these stars is the late Herbig~Be star, HD~100546 (B9.5Ve,
103pc; \citet{van_den_ancker_1}).  Its circumstellar environment was first
resolved in scattered light by \citet{pantin_1}, revealing the
presence  of   a  disk   with  an  inclination  of  $\sim$50$^\circ$.
Coronagraphic imaging showed that the star is surrounded
by a large scale envelope ($\sim$1000~AU) and a disk extending up to 515~AU whose 
brightness profile is asymmetric  \citep{augereau_1,grady_1}.  Its
position in the  HR diagram gives an estimated  age larger than 10~Myr,
supported by the presence  of large fractions of crystalline silicates
that suggest an evolved disk \citep{bouwman_1}.

The SED of the system has been intensively studied. Compared to other
HerbigAeBe  stars, its specificity  is to show a strong mid-infrared
excess,  peaking at  40~$\mu$m,  compared to  a  rather weak  2-8~$\mu$m
emission.   To account  for  the  deficit of  NIR  excess, \citet{bouwman_1}  suggested the  presence of  an inner  cavity  (empty region)
within  10~AU,  while  fitting  the  SED  with  three  optically  thin
concentric shells of different grain populations. Besides, support for 
a region of low  density  in  dust grains  is  provided  by the  low
circumstellar  polarization in  comparison with  younger  Herbig stars
\citep{clarke_1}.  Spectroscopic observations also agree with this 
conclusion. STIS  observations of H$_{2}$ emission revealed the
presence    of     a    central    cavity     extending    to    13~AU
\citep{grady_2}. Similarly, \citet{acke_1} concluded to the presence
of a gap distant of 6.5~AU from the star by studying the 6300~\AA~[OI]
line. The last authors speculated that this gap (or secondary wall) is
induced   by  a  very   low-mass   stellar  or   planetary  companion
(20~M$_{\rm{Jup}}$).   This could  be  the explanation  for the  pericenter
asymmetry observed  in the reflection  nebulosity within the  cavity by
\citet{grady_2} that could else not be  due to chemistry effects
(asumming  that the  density  distribution is  axisymmetric).
The recent CO observations \citep{brittain_1, van_der_plas_1} reported an extended emission with a large inner hole inside
$\sim$11-13~AU,  confirming the  status of  \textit{transitional} disk
given to HD~100546.

Nevertheless, the measured NIR excess as well as the presence of gas located
from 0.5 to 60~AU  \citep{acke_1,martin_zaidi_1}, imply that the inner
AUs are not completely empty.  In the mid-infrared wavelength range, the disk is resolved with
mid-infrared spectro-interferometry \citep{leinert_1} and nulling
interferometry showed that the radial temperature law in the few inner
AUs (at 10~$\mu$m and 20~$\mu$m) is not compatible with a continuously flared disk \citep{liu_1}.  
\citet{vinkovic_1} successfully reproduced  the NIR SED  using an
optically thick disk with an inner radius at 0.45~AU, surrounded by an
optically thin dusty enveloppe. All these results point toward structural
differences between the inner and outer regions of the HD~100546 disk.  

Constraining  the spatial  structure  and physical  conditions in  the
innermost disk is therefore a crucial issue. In particular, the exact position and geometry 
of the inner rim remain poorly constrained. NIR interferometric observations
can  achieve  the  spatial  resolution  needed  to  resolve  this  hot
circumstellar material.  The
combination   of  angular  and   spectral  resolution provided by
AMBER at the ESO/VLTI allow us to directly resolve the circumstellar
material located in the first AU of the disk and to address the  question of its
geometry.  In this paper, our goal is to revisit the disk geometry with a special emphasis 
on its inner AU. The  radiative transfer models we have calculated aim
at reproducing both the interferometric and photometric observations. 

The paper is organized as follow: in Sect.~2, we first describe the
observations and the data reduction procedure; in Sect.~3 and 4, we
present  the interferometric  results followed  by the  description of
our disk models. Finally, we discuss our results in Sect.~5.

\section{Observations and data processing}
\label{sec:amber}

\begin{table}[t]
\centering
\caption{Log of the observations, obtained  on June 24, 2006, with the
  UT3-UT4 baseline} 
\label{tab:obs}
\begin{tabular}{c c c c c c }
 \hline
 \hline
Observation  & Projected & Position & Spectral & Resolution \\
 number & length (m)& angle ($^\circ$)& range ($\mu$m) & (mas)\\
 \hline 
% 2.06-2.46 & U1-U3 & 82.4 & 39.8 & 6.2 \\
\# 1 & 59.8 & 99.5 & 2.06-2.46 & 7.1-8.5\\
% ~ & U1-U4 & 122.6 & 65.5 & 4.2 \\
\# 2 & 60.4 & 104.5 & '' & 7.0-8.4\\

 \hline
 \hline    
\end{tabular}
\end{table}

\begin{table}[t]
\begin{center}
 \caption{Stellar and calibrator properties}
 \label{tab:calparam}
 \begin{tabular}{c c c c c c}
   \hline
   \hline
   Star & V & K & H & Sp. Type & Diameter \\
          &     &    &    &                 &   [mas]    \\
   \hline 
   HD100546 & 6.7 & 5.4& 5.9& B9Vne & --- \\
   HD101531 & 7.3 & 4.2 &4.3 & K1III& 0.6$\pm$0.1 \\
   \hline
   \hline    
 \end{tabular}
\end{center}
\end{table}

HD~100546  was  observed at  the  Very  Large Telescope  Interferometer
(VLTI; \citet{haguenauer_1}) with the AMBER instrument on 24 April 2006.  
AMBER simultaneously combines three beams in the
near  infrared H  and K  bands  using spatial  filtering and  spectral
dispersion  \citep{petrov_1}.   It  therefore provides  spectrally
dispersed visibilities and closure phases.  In the 
following, we  present observations that were obtained  with three 8-m
unit  telescopes  (UT1,  UT3,  UT4)  and AMBER  in  its  low  spectral
resolution mode (R$\sim$30).  In 
addition to HD~100546, a calibrator target of known diameter (HD~101531)
was observed. 
%in order to derive systematic effects on 
%the interferometric observables due to the instrument. 
Its properties can be found  in Table~\ref{tab:calparam}. The  observations were achieved
under good  weather conditions  (seeing of $\sim$0.8''  and coherence
time of $\sim$5ms) and without  fringe tracker.   The data  consist of
respectively, 8 and 5 exposures of 1000 frames each, for HD100546
and its calibrator.

The  data  reduction   was  performed  following  standard  procedures
described in \citet{tatulli_1}, using the \texttt{amdlib} package,
release 2.99, and the \texttt{yorick} 
interface,  provided by  the Jean-Marie  Mariotti Center  (JMMC).  Raw
spectrally dispersed visibilities and closure phases are extracted for
all the frames of each  observing file. A selection process among them
is  necessary  to  avoid   the  effects  of  instrumental  jitter  and
unsatisfactory  light  injection.  Various selection  procedures  were
applied, including selection based on the fringe signal-to-noise ratio (SNR) and on
the fringe optical path difference (OPD; \textit{i.e.} piston). Although no
difference  in  absolute  values  was  found between  them,  the  best
accuracy was obtained with a selection of 20\% of the best exposures based on
the fringe SNR.  This criteria takes into account both the flux level and
the  fringe  contrast (or  visibility).   Among  all the  measurements
obtained on the three baselines, only the ones corresponding to the shortest
baseline (UT3-UT4) in K-band, led to SNR above 1.5.   Consequentely,  all
H-band measurements, K-band data corresponding to UT1-UT3 and UT1-UT4 baselines,
and all  the  closure  phases,  were  rejected.   The
remaining files have SNR ranging from 2.3 to 5.8.  The same selection was applied to both the
scientific star and calibrator measurements in order to guarantee an uniform data reduction.  The attenuation in fringe contrast due to
additional piston with respect to the zero OPD was corrected following
\citet{tatulli_1}.  Finally, calibration of the AMBER+VLTI
instrumental transfer function was done using measurements of HD~101531,
after  correcting for  its  diameter.  The  errors  on the  calibrated
spectral  visibilities  include the  statistical  error obtained  when
averaging individual frames, the error on the 
diameter of the calibration star, and the dispersion of the transfer
function along the calibrator measurements. The last element is the
one that dominates the error budget.  After data processing, the final
data  set consists  in 2  measurements on  UT3-UT4 baseline,  that are
spectrally dispersed in 13 K-band spectral channels.  Therefore, in total, we obtained 26
visibilities in the [2.06-2.46]~$\mu$m wavelength range.  A summary can
be found in Table~\ref{tab:obs}.

\begin{figure*}[t]
 \centering
 \begin{tabular}{cc}
   \includegraphics[width=0.39\textwidth]{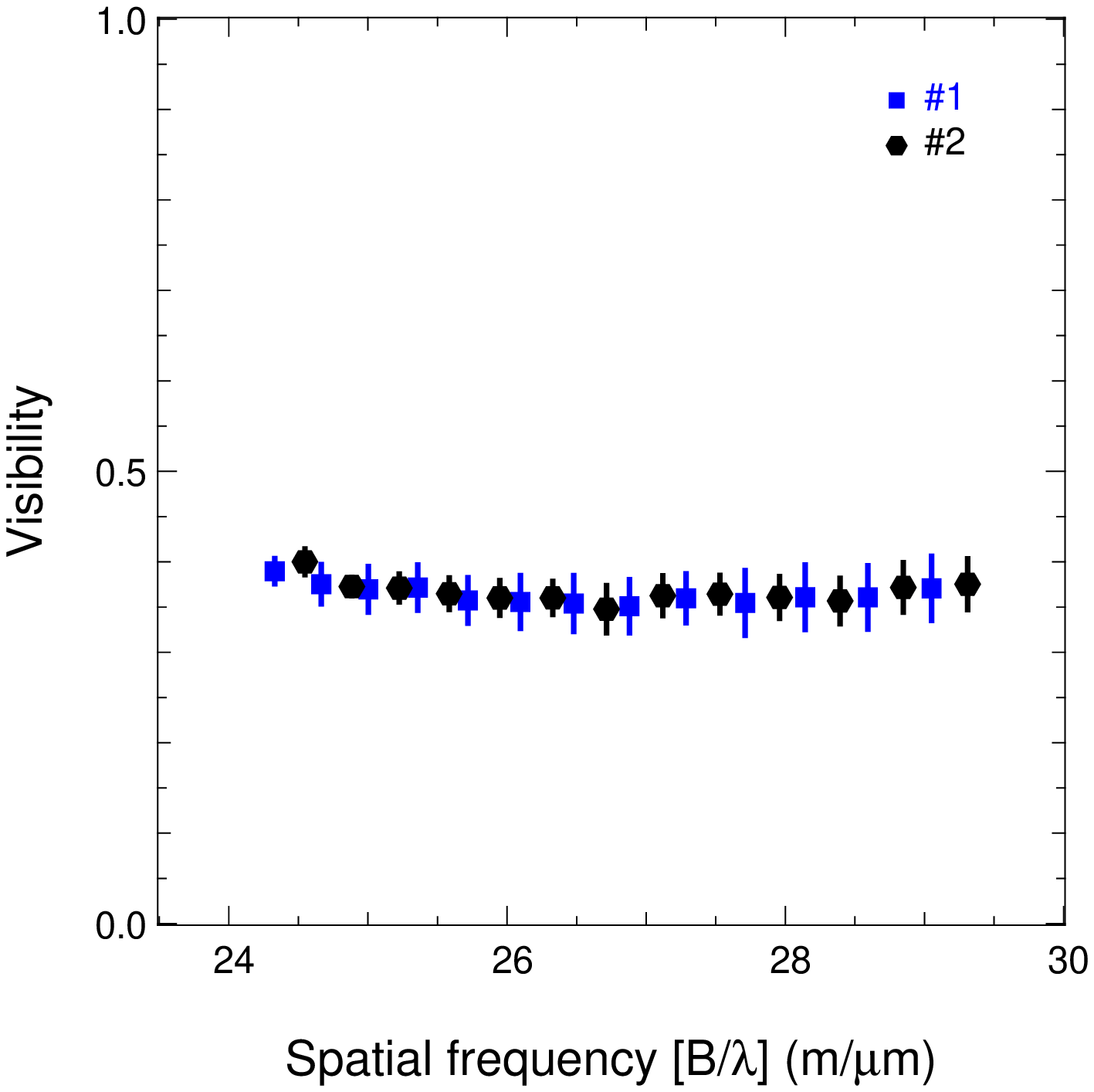} &
   \includegraphics[width=0.38\textwidth]{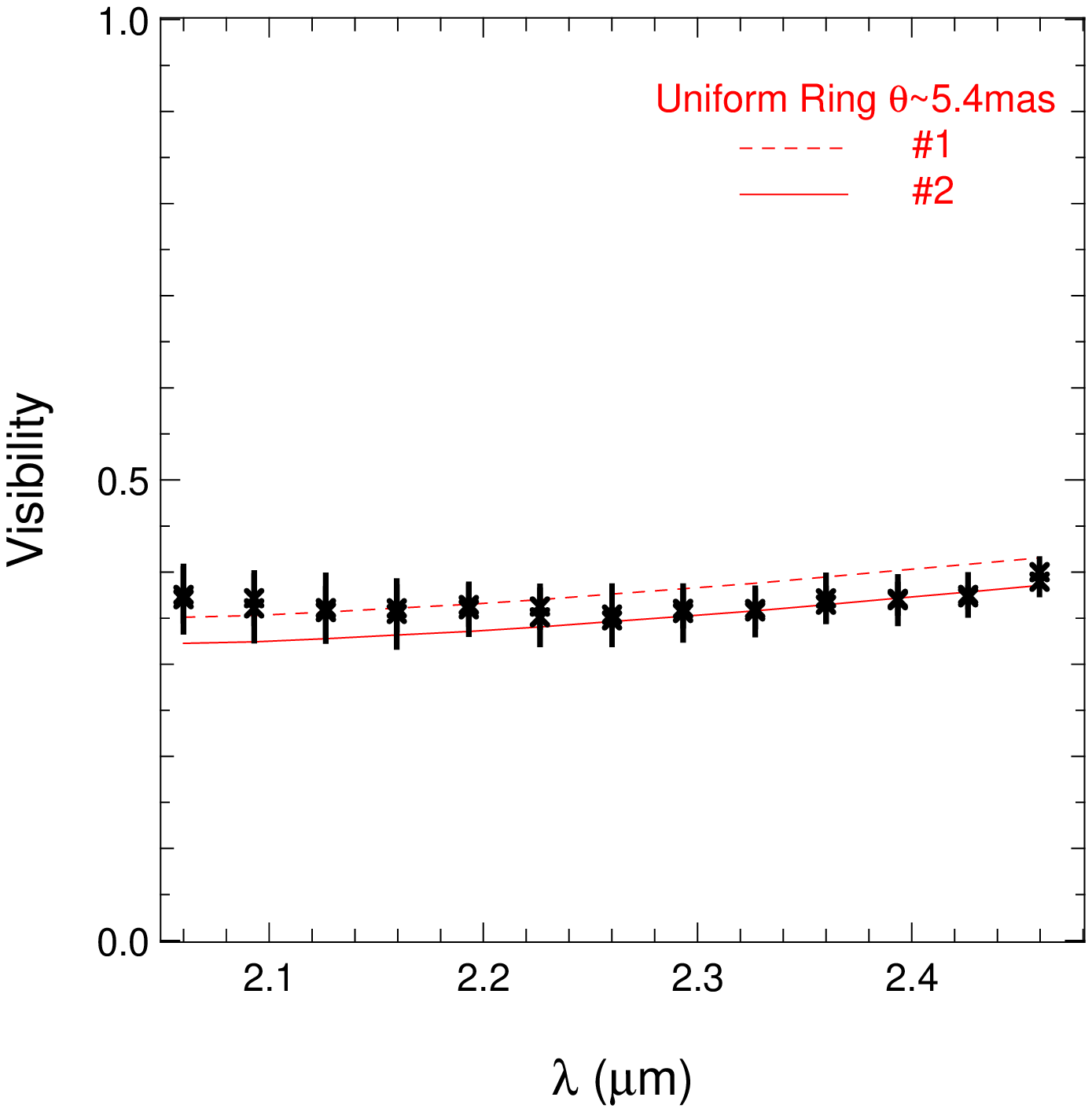} 
 \end{tabular}
 \caption{\label{fig:vis} Left:  measured visibilities plotted against
   spatial frequency.   The two different measurements  (\#1, \#2) are
 shown with distinct symbols. 
 Right:   the  visibilities  (black   crosses)  are   shown  against
   wavelength and are compared to the predictions of a ring of 
   uniform  brightness.  The red  full  and  dashed  lines give  these
   predictions  for  the   two  measurements  (\textit{i.e.}  at  two
   distinct projected baselines).}
\end{figure*}

\section{Results}

\subsection{Visibilities}
\label{sec:vis}

Fig.~\ref{fig:vis}, left, shows the measured visibilities after data
processing. They are plotted against the spatial frequency, B/$\lambda$, with B the
projected  baseline  length  and   $\lambda$,  the  wavelength  of  the
observations.  The spatial frequency  is the inverse of the resolution
achieved by the interferometer  and takes into account the wavelength,
the projected baseline, and consequentely, the 
direction on the  sky.  It can be expressed in  unity [m/$\mu$m] or in
cycles per  milli-arcsecond. It  can be immediately  seen that
  the circumstellar  matter around HD~100546 is resolved  at the level
  of a few milli-arcseconds, because the visibility is well below~1.

To derive  basic characteristics of the NIR  emission, one can
  consider a simple geometrical model  of a ring of uniform brightness
  that roughly accounts for hot dust located in the inner rim of a 
dusty disk.  The model visibilities V$_{\rm{mod}}$ can be written~:
$$  V_{\rm{mod}}  =  \frac{  V_{*}+  V_{\rm{ex}}*F_{\rm{ex}}/F_{*}}{1+
  F_{\rm{ex}}/F_{*}}$$ \noindent with V$_{*}$=1 (unresolved central star) and
F$_{\rm{ex}}$/F$_{*}$, the  ratios of the NIR excess  to the stellar
  flux.  This  ratio, $\sim$2.2,  is estimated at  each AMBER
  wavelength, using a Kurucz model for the central star 
emission at T=10500~K. With this model, the extension of the K-band emission is found to be $\sim$5.4~mas
(\textit{i.e.} a radius of $\sim$0.28~AU) at d$=$103~pc.  
The observed chromaticity  of the visibilities can be  compared to the
predictions  of  this  model.   Fig.~\ref{fig:vis}, right,  shows  the
  spectral visibilities against wavelength. 
The visibility curve is almost flat with wavelength, while the uniform
brightness  ring  predicts  a  slightly  steeper  slope.

HD~100456 was also observed by  \citet{leinert_1} in the mid-infrared
with  the  MIDI instrument  at  VLTI.   These  observations led  to  a
single-dish  image  at  8.7~$\mu$m  and  to  visibility
measurements (at $74$~m baseline length).  In Sec.~4, we use this image
and the corresponding 8.7~$\mu$m visibility. 

\subsection{Photometry}
We retrieve from the  litterature photometric measurements from the UV
to the millimetric wavelengths to produce a complete SED (2MASS, IRAS,
ISO;   \citet{ardila_1,henning_1,grady_1}).
Moreover,  because  young  stars   are  highly  variable  in  the  NIR
wavelengths  \citep{sitko_1},  and  since   modelling  of  the
observations relies on these 
photometric values, it is useful to provide an estimate of the
NIR emission directly from the interferometric measurements. We were able to extract the K band photometry from the
three  AMBER photometric  channels.  Following  \citet{acke_2}, for
each file, we summed the flux in each channel over all frame and 
calibrated this value with the total flux in the corresponding 
channel for  the calibrator. We  applied a photospheric model  for the
calibrator star in order to convert it in flux units and 
obtain 5.33$\pm$0.08 as the K band magnitude, consistent with
the 2MASS values.

\begin{figure*}
\begin{center}
\begin{tabular}{cc}
\includegraphics[width=1.37\columnwidth]{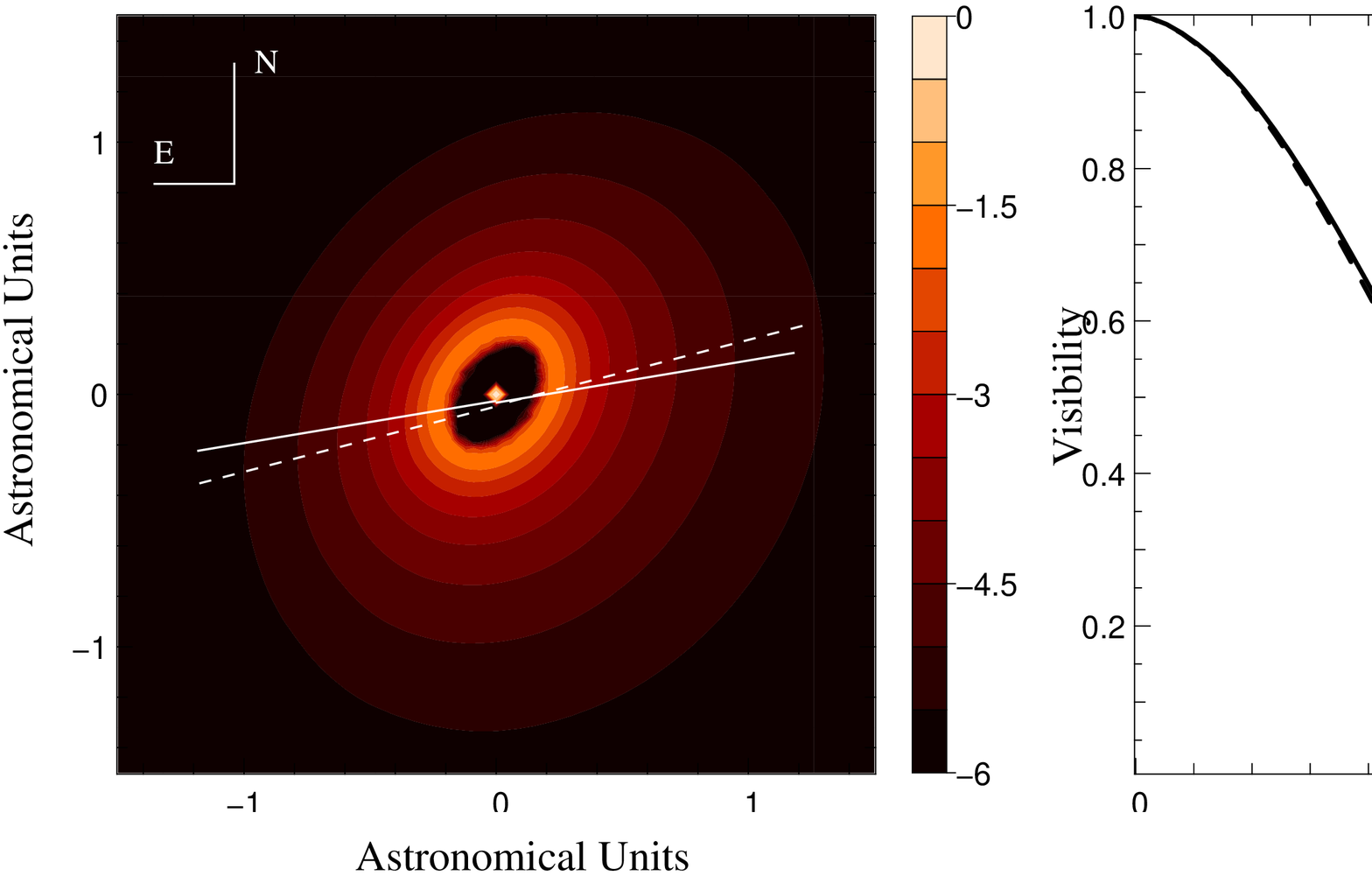} & 
\includegraphics[width=0.59\columnwidth]{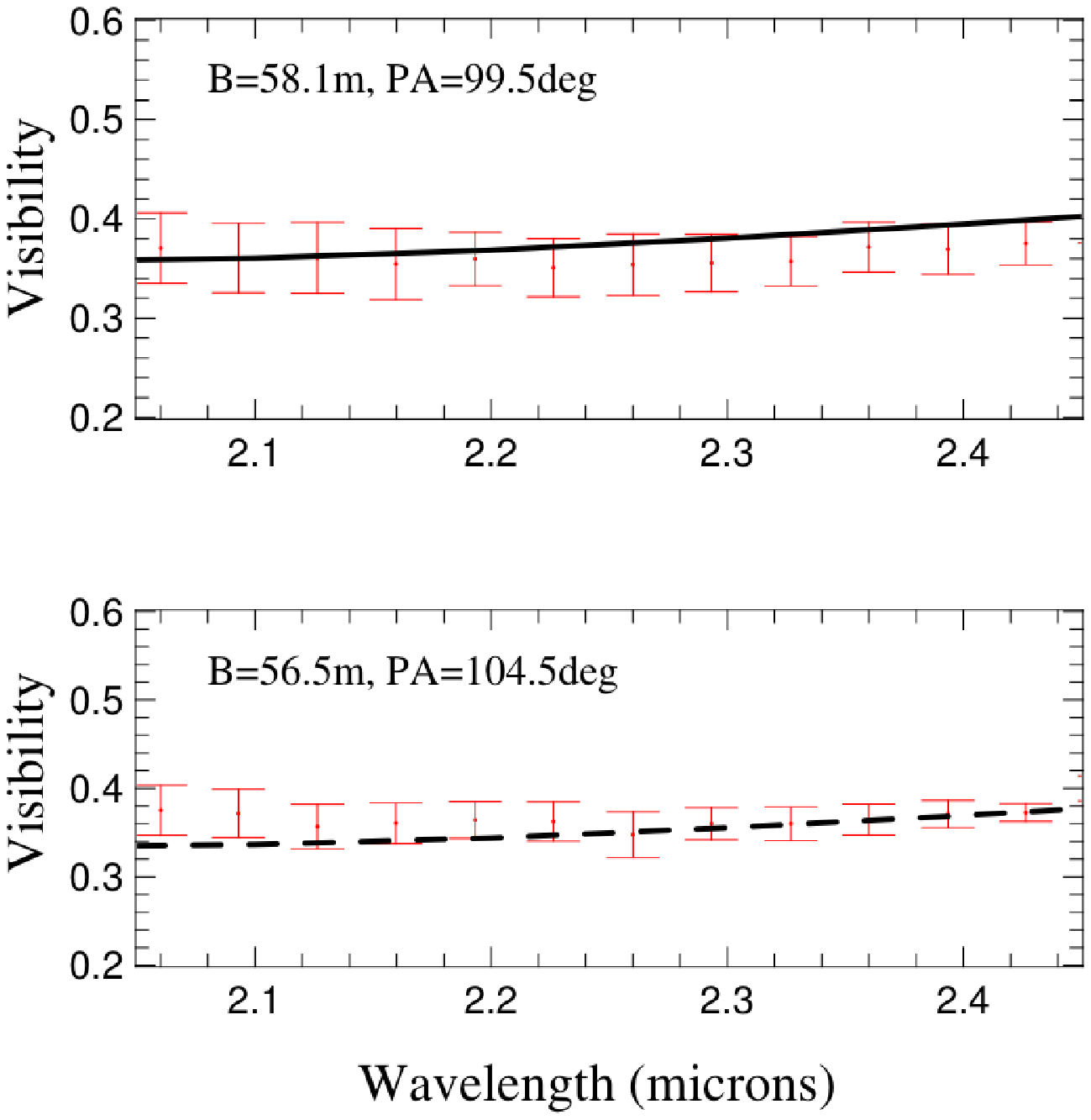}
\end{tabular}
\caption{Left: K-band  model image  of HD100546. The  full and
    dashed  lines indicate  the  position angles  of the  observations
  ($PA=99.5^{\circ}$ and $PA=104.5^{\circ}$, respectively). 
  Middle: model  visibilities,  as a function
  of baseline  length.  The broad  band K-band AMBER  visibilities are
  overplotted.  Right: spectrally dispersed AMBER visibilities versus
wavelength, for  both  measurements   (upper  and   lower  pannels
  respectively).  The spectral profiles  of the model visibilities, as
  derived from our best model, are added. } 
\label{fig:VIS}
\end{center}
\end{figure*}

\section{Modelling}
Since it has been long established that HD~100546 is surrounded by a disk
\citep{augereau_1}, in  this  section,  we  use disk  models  to
describe  its  circumstellar  environment.   It  is  not  possible  to
unambiguously estimate the disk inclination and position angle from 
a single-baseline interferometric measurement, therefore, we adopt the
previously published values of 42$^\circ$ and 145$^\circ$, respectively
\citep{ardila_1}.  Besides,  we consider  a total disk  mass of
5x10$^{-4}$  M$_{\odot}$,  inferred   by  1.3~mm  imaging  \citep{henning_1}.

\subsection{The radiative transfer code MCFOST}
\label{secmcfost}
We used the Monte Carlo-based 3-D radiative transfer code, MCFOST, to
produce disk images and SED \citep{pinte_1}.  The temperature
distribution is calculated using  the immediate re-emission concept of
\citet{bjorkman_1} but  with a continuous deposition of energy to
estimate the mean intensity \citep{lucy_1}. In this paper, we used a
cylindrical grid,  with an adaptive mesh  at the inner  edge (based on
the opacity gradient) so as to properly sample the inner radius of the
disk.   The radiation field and temperature
structure estimated by  the Monte Carlo runs are  used to produce SEDs
by  calculating the formal  solution of  the radiative transfer equation along
rays  (ray-tracing  method),  as  well as  images.   Visibilities  are
computed by taking the Fourier transform of the latter, at the specific spatial frequencies sampled
by the  observations.  The  code has been  extensively tested  and its
performances compared to other radiative transfer codes \citep{pinte_2}.

We assume that the disk is constituted of silicate dust grains
  only  \citep{pollack_1}, and  is in  hydrostatic  equilibrium.  Its
structure is parametrized by power laws, so that the surface density $\Sigma$, and
  the  scale height~H  can be  expressed  as varying  with radius,  r,
  following~: 
$$\Sigma(\rm{r}) = \Sigma_{100\rm{AU}} \left(\frac{\rm{r}}{100\rm{AU}}\right)^{q}$$ 
$$\rm{H}(r) = \rm{H}_{100\rm{AU}}  \left(\frac{r}{100\rm{AU}}\right)^{\beta}$$ 

$\beta$  is the  flaring  exponent. $\Sigma_{100\rm{AU}}$  and
$\rm{H}_{100\rm{AU}}$ are the values of the surface density and
the height  scale at a  radius of 100~AU.   The dust
grain sizes, a, are distributed following a power law a$^{-3.5}$ (ISM), with
a varying between the  minimum and maximum sizes, a$_{\rm{min}}$ and
a$_{\rm{max}}$  respectively.   In addition,  we  fix  the total  dust
mass (M$_{\rm{dust}}$) of the disk, that is delimited by its
inner and outer radii (R$_{\rm{in}}$ and R$_{\rm{out}}$).

The disk of HD~100546 is well  known and its geometry and content have
been          estimated           by          several          authors
\citep{grady_1,bouwman_1,brittain_1}. We will therefore
not  provide  a  detailed  re-analysis  of  all  the  disk  structural
parameters. To the  contrary, using in spirit the  results of previous
studies, we  will construct  a disk model  that also includes  a large
gap.  However,  with the  help  of  interferometric measurements,  and
complete radiative transfer including  scattering, we will revisit the
morphology and mass content of the inner disk, as well as the position
of the gap, in a global picture.

\subsection{The disk model}
\begin{figure*}
\begin{center}
\begin{tabular}{cc}
\includegraphics[width=0.82\columnwidth]{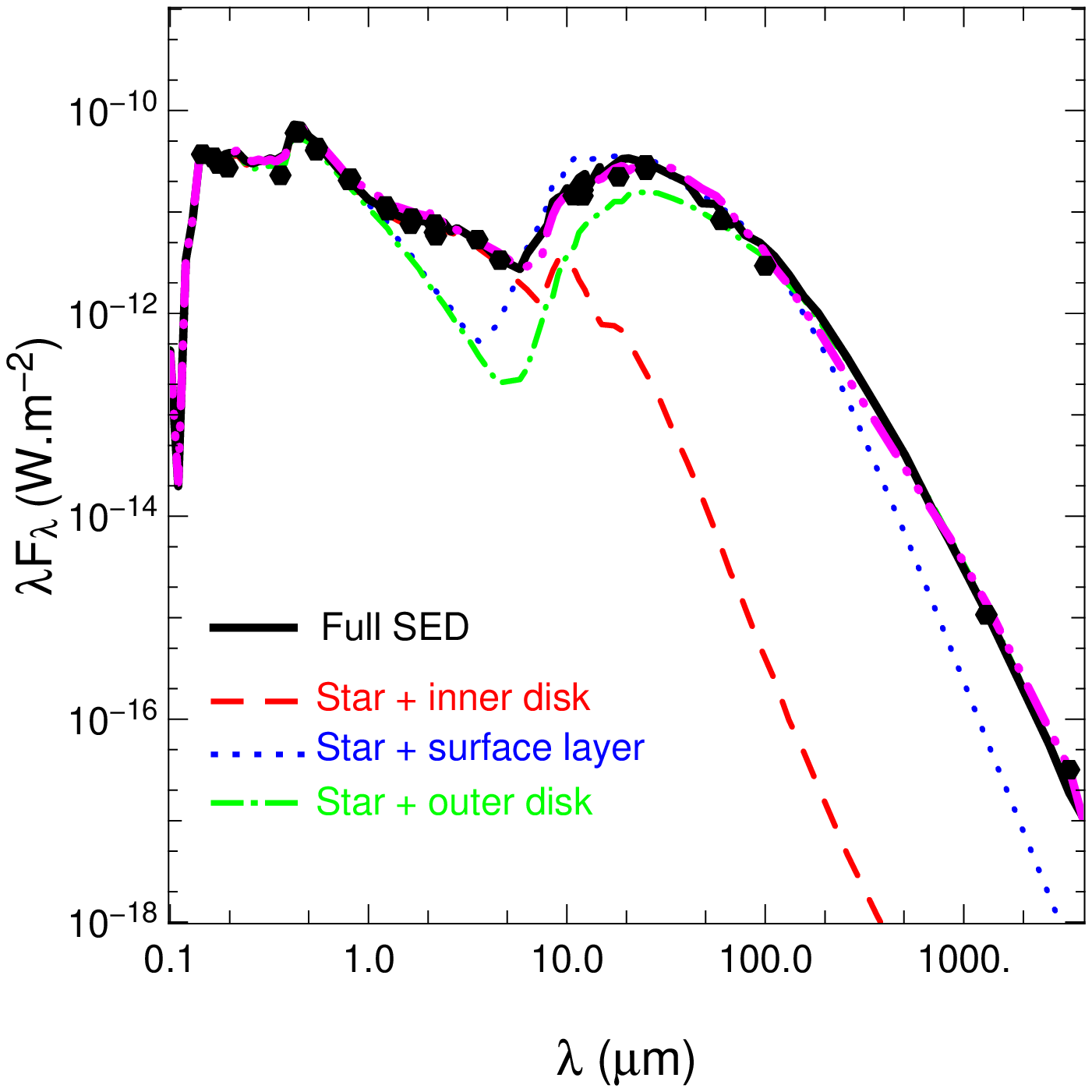}
& 
\includegraphics[width=0.82\columnwidth]{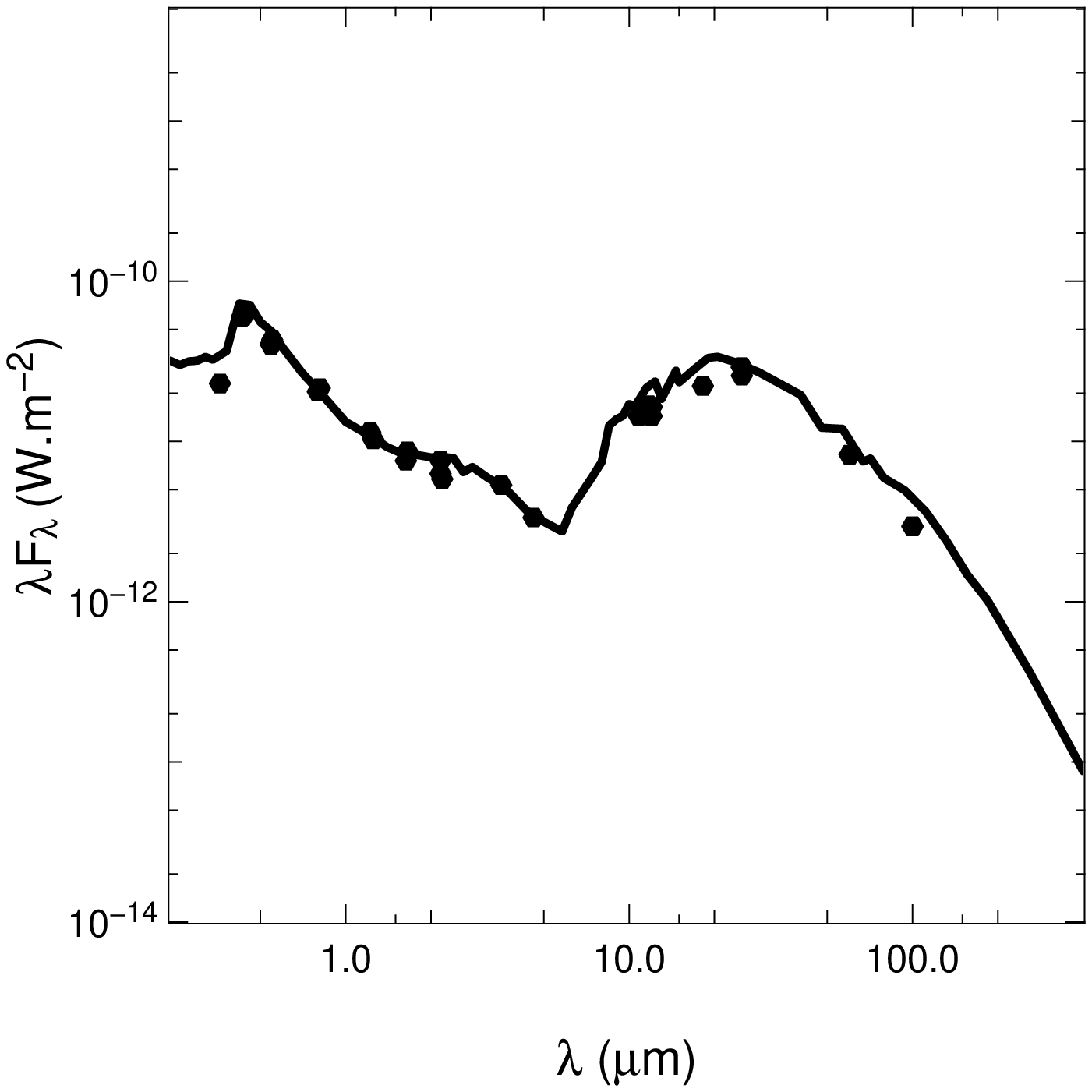} 
\end{tabular}
\caption{Left: measured SED (circles) for HD~100546, compared to
  the predictions of our best model (black full line).  The individual
  contributions of each disk zone to the 
  SED are shown (inner disk, red dashed curve; surface layer, blue dotted
curve; outer disk, green dash-dotted line). A similar fit to the SED
  can  be  obtained with  a  different  disk model  (magenta
  line). Right: zoom in the near and mid-infrared wavelengths.}  
\label{fig:SEDmodele1et2}
\end{center}
\end{figure*}

\begin{figure*}
\begin{center}
\includegraphics[width=2.0\columnwidth]{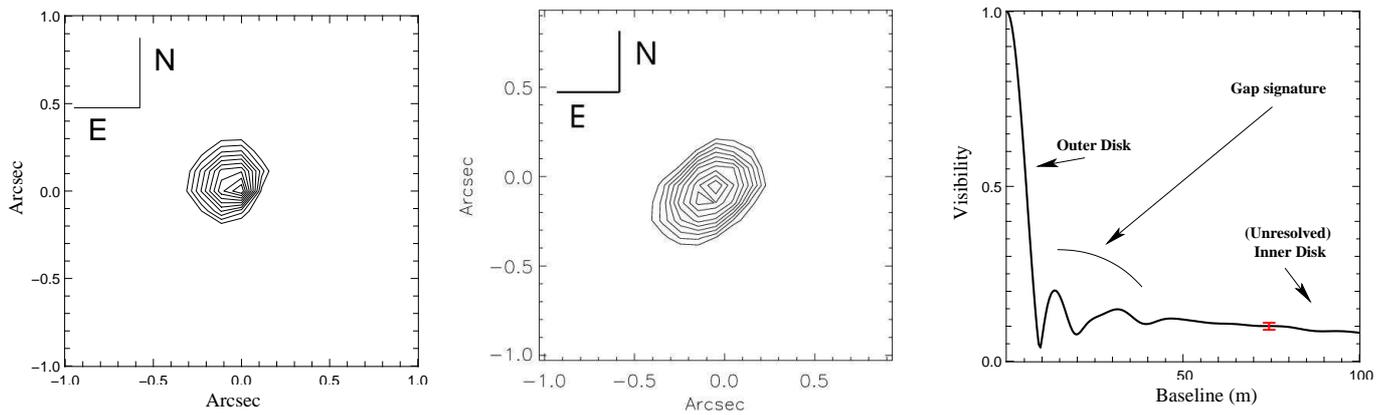}
\caption{Left:  8.7~$\mu$m continuum image  of HD~100546  predicted by
  our best model and oriented along a $145^{\circ}$ position angle. Middle:
  8.7~$\mu$m  continuum  image  obtained  with  a  single  UT  at  VLT
  \citep{leinert_1}.  For both images, the contour levels are linearly
  spaced from $100\%$ to $25\%$ as the fourth root of the image
  brightness. 
  Right: visibility curve calculated for our best model (solid line) at
  8.7~$\mu$m. The MIDI measurement  obtained by \citet{leinert_1} is
  shown with the error bar. }
\label{fig:continuum8.7}
\end{center}
\end{figure*}

The SED constrains the mass-temperature  law for the dust. The near-infrared emission  in excess of the  photosphere is
due  to hot material  at $\sim$1500~K,  and must  result from  a small
amount  of  dust in  an  optically  thin  inner  disk. The
large mid and far-infrared excesses are due to dust at a temperature of $\sim$200~K, while large and cold
grains  in the  mid-plane of  the disk  must be  responsible  for the
emission  at longer  wavelengths.  The  SED gives  integrated flux
informations but provides little morphological informations on the dust spatial
distribution.  \\
In  this  section,   we  propose  a  disk  model  that
successfully accounts for the emission  at all wavelengths, as well as
for   the  AMBER  and   MIDI  visibilities   (see  Fig.~\ref{fig:VIS},
\ref{fig:SEDmodele1et2} and \ref{fig:continuum8.7}).  However, our model does
not  attempt  to  match  mineralogical  features. As  pointed  out  by
\citet{bouwman_1}, continuous flared disk models can not reproduced the SED as
they produce either too much 
or  too  little  flux   in  the  near  and  mid-infrared  wavelengths,
respectively.  The  model includes an  optically thin
inner  disk,  a gap  and  an optically  thick  outer  disk. The  model
parameters can  be found  in Table~\ref{tab:modelparam}, including  for each
zone, the mass (M$_{\rm{dust}}$), the inner and outer radii
(R$_{\rm{in}}$, R$_{\rm{out}}$ respectively), the scale height at
100~AU (H$_{\rm{100AU}}$), the flaring and surface density exponents
($\beta$, q, respectively) as well as the minimum and maximum grain
sizes (a$_{\rm{min}}$, a$_{\rm{max}}$, respectively).\\

\begin{figure*}[htbp]
\begin{center}
\includegraphics[width=1.7\columnwidth]{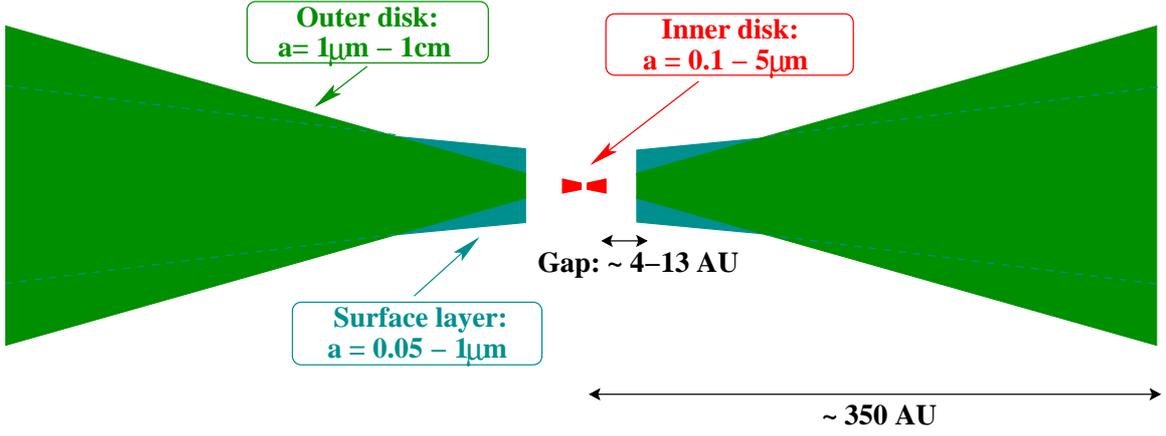}\\
\caption{Schematic  view  of  our   best  disk  model,  that  includes
  different zones: an inner  disk with micron-sized silicate grains, a
  gap, a massive outer disk with small grains in an upper layer.} 
\label{fig:diskfig}
\end{center}
\end{figure*}

\textbf{The inner disk:}
The AMBER/VLTI measurements can locate  the bulk of the emission in the
K-band at $\sim$0.25-0.30~AU. This puts a strong constraint on the
location of the disk inner edge and consequentely on
the silicate dust grains sizes.  The tenuous inner
disk has a total dust mass of 3.10$^{-10}$ M$_{\odot}$ and starts at a
radius  of  $\sim$0.26~AU, a  distance  inferred  from sublimation  of
micron-sized silicate grains (specifically, with sizes ranging from 0.1 to 5~$\mu$m) at
$\sim$1500~K.  Its outer  radius is not well constrained  by the AMBER
observations because it  is fully resolved, and we  fix it at
$\sim$4~AU.  Here, all the model calculations are done assuming that the dust
grains  are  at the  same  equilibrium  temperature  with the  stellar
radiation  field which  may not  be exactly  the case  in  the tenuous
optically thin  inner regions.  Although a  more detailed calculation,
that requires determination of the dust-to-gas ratio is
beyond the goal of this paper, we stress that there may not be a
specific inner  radius but  a range of  inner radii, depending on the grain sizes. \\

\textbf{The gap:} After computing numbers of models, we confirm the
need for a very low  density region (\textit{e.g.} a gap) to reproduce
the lack of emission in the 3-10~$\mu$m wavelength range. The
MIDI/VLTI measurements provide a  complementary information on  the spatial
distribution  of  the  dust  in  the disk.  At  the  achieved  angular
resolution  at  $\sim$8.7~$\mu$m  (24~mas  \textit{i.e.}  2.4~AU),  the
innermost disk is not resolved but the outer disk is fully resolved. 
The 8.7~$\mu$m observations therefore bring a strong constraint on the
flux ratio, between these two elements of the disk.  
In the model, the gap starts at 4~AU until 13~AU, a radius consistent 
with spectroscopic results \citep{grady_2, brittain_1}.\\

\textbf{The outer disk and its surface layers:} the model massive outer disk forms at 13~AU and extends up to 350~AU.  It contains most of the
disk mass  and holds large dust  grains (from 1~$\mu$m  to 10~mm).  To
account for the mid and far-infrared excesses, we include a population of small grains
(\textit{i.e.} 0.05  to 1~$\mu$m)  at the surface  of the  outer disk,
at distances of a few tens of AUs (here, from 13 to 50~AU; see
 Table~\ref{tab:modelparam}). The
latter are not precisely constrained by the
observations, as the  only requirement is to have  a specific amount
of small grains emitting at a few tens of AUs.  \\
The  entire  set  of  parameters  used  in  this  model  is  given  in
Table~\ref{tab:modelparam}, and a schematic view of the disk model is 
provided in Fig.~\ref{fig:diskfig}.  In Fig.~\ref{fig:SEDmodele1et2}, we
report the contributions from each part of the disk to the total flux. The
inner disk (red dashed line) produces enough  emission to fit
the NIR excess while the small grains in the surface layers (blue
dotted line) produce the major  contribution to the large mid and far
infrared  excess.  Finally,  the large  grains in  the outer  disk are
responsible for the flux at long wavelength (green dash-dotted line). 
The predictions  of the  entire model are  in good agreement  with the
observations (full black line).\\

\vskip 0.1cm
For this  model, images  and visibilities were  computed at  the AMBER
wavelengths.  Figure~\ref{fig:VIS} (middle) shows the visibility profile as
function  of the  projected baseline  length.  This  model succesfully
reproduces the  interferometric observations  in addition to  the SED:
the spatial extent of its K-band emission and its inner structure are
in good  agreement with the data.   The chromaticity in  the K-band is
also well reproduced with the radial temperature profile of the disk model (Fig.~\ref{fig:VIS}, right).  
In the mid-infrared, we focus our modelling on the continuum emission
at $8.7$~$\mu$m and do not  attempt to reproduce any spectral feature.
Figure~\ref{fig:continuum8.7} (left) gives the model image at
$8.7~\mu$m that clearly shows an elongated disk structure, in agreement
with      the       one      obtained      by     \citet{leinert_1}
(Fig.~\ref{fig:continuum8.7}, middle).  From a simple 2-D fit, we 
obtain $\sim$21~AU and $\sim$14~AU as the disk major and minor axis,
respectively,   in  good   agreement  with   the  values   derived  by
Leinert~et~al.~2004  ($29 \pm  4$~AU  and  $19\pm 9$~AU
respectively)   and   compatible  with   the   results  from   nulling
interferometry  ($24$~AU; \citet{liu_1}).  
Figure~\ref{fig:continuum8.7}, right, shows the model visibilities
at 8.7~$\mu$m,  compared to the MIDI measurement.   At long baselines,
the outer disk is entirely resolved and one has 
access to  the unresolved flux contribution  of the inner  disk to the
total 8.7~$\mu$m-flux.  At a 74~m-baseline, our model
can  well account  for this  flux ratio.   Our  model predicts
  that, at short baselines ($\leq$10~m), the outer disk is resolved as 
the steep decrease of visibility attests, while for intermediate
baselines, the  sharp edge  of the gap  produces a clear  signature. 
Mid-infrared interferometry therefore appears well suited to probe the
presence of a gap, as the  strong gap signature can easily be detected
using moderate VLTI baselines.

\begin{table}[tdp]
\caption{Parameters of our best model.  } 
\begin{center}
\begin{tabular}{l l || c c c }
\hline
\hline
 & & inner disk & surface layer & outer disk \\
\hline
M$_{\rm{dust}}$ & [M$_{\odot}$] & 3e-10 & 6e-5 & 5e-4\\
R$_{\rm{in}}$ & [AU] & 0.26 & 13 & 13 \\
R$_{\rm{out}}$ & [AU] & 4 & 50 & 350\\
H$_{\rm{100AU}}$ & [AU] & 6 & 12 & 12\\
 $\beta$ & & 1 & 0.5 & 1.125\\
q & & -1 & -1 & -1\\ 
a$_{\rm{min}}$ & [$\mu$m] & 0.1 & 0.05 & 1\\
a$_{\rm{max}}$ & [$\mu$m] & 5 & 1 & 10000\\
%n & -3.5 & -3.5 & -3.5\\
\hline
\hline
\end{tabular}
\end{center}
\label{tab:modelparam}
\end{table}

\begin{table}[tb]
\centering
\caption{Contributions from the different types of emission to the total flux.}
\label{tab:flux}
\begin{tabular}{c ||c c c c}
 \hline
 \hline
Model with & F$_{\rm{dir,st}}$/F$_{\rm{tot}}$ & F$_{\rm{sc,st}}$/F$_{\rm{tot}}$ &
 F$_{\rm{dir,th}}$/F$_{\rm{tot}}$ & F$_{\rm{sc,th}}$/F$_{\rm{tot}}$\\
 \hline 
R$_{\rm{in}}$=0.26~AU & 19.6 & 4.1 & 38.0 & 38.3 \\
R$_{\rm{in}}$=0.40~AU & 16.7 & 1.8 & 64.4 & 17.1 \\
 \hline
 \hline    
\end{tabular}
\end{table}

\section{Discussion}
In this paper, we propose a model of the disk surrounding
HD~100546  that  successfully  reproduces   the  SED,  AMBER  and  MIDI
interferometric observations.   We note that equally good  fits to the
SED can be obtained with other 
models,  similar in spirit  to the  one we  propose (\textit{i.e.}
with three distinct disk regions) but with different spatial
distribution. In particular, \citet{bouwman_1} suggested a solution
to reproduce the SED, that includes a population of small grains distributed from 0.3 to 9.8~AU, a more massive part with small grains
from 9.8  to 43~AU, and a population  of large dust grains  from 28 to
380~AU, with a grain size distribution n(a) $\propto$ a$^{-2}$.  Different
species  are used  to study  the mineralogy of  the dust,
including refractory grains, such as  iron.  Using the same  grain sizes
distribution, but for silicate only, we could produce a good fit to the
SED (see Fig.~\ref{fig:SEDmodele1et2}, left, magenta curve).  To account for the NIR emission, the small grains are nevertheless required to be
hotter than the tabulated evaporation temperature for silicate (\textit{i.e.} 
2100~K instead of 1500~K). Besides, they induce a slightly different location for
the bulk of the  K-band emission (\textit{i.e.}  0.4~AU) which
  does not  match the AMBER  visibilities.  We therefore  confirm here
  that spatially resolved informations  are needed to speculate on the
  disk structure  and properties, as  fitting the SED alone  is highly
  ambiguous.\\ 

\textbf{The importance of scattering:} An interesting side effect is the amount of scattered light
induced in different models.  At each wavelength, the total emission
(F$_{\rm{tot}}$)  is  constituted  of  4 distinct  contributions:  the
stellar light  that directly reaches  the observer (F$_{\rm{dir,st}}$);
the stellar  light that is  scattered by the  disk (F$_{\rm{sc,st}}$);
the  direct thermal   emission    from   dust   grains   (F$_{\rm{dir,th}}$;
\textit{i.e.}  absorbed and then  re-emitted photons) and finally, the
scattered thermal light (F$_{\rm{sc,th}}$; \textit{i.e.}  absorbed,
re-emitted and then  scattered photons).  Micron-sized silicate grains
located at  0.26~AU are good scatterers, since  their sizes are
of the order of the AMBER wavelengths.  In our model, the scattered thermal light contributes to
$\sim$38\%  of the  total K-band  emission, about as much  as the  direct
thermal flux.   On the  contrary, the model  with small grains  in the
inner disk and an inner radius at 0.4~AU, predicts a dominant contribution from direct
thermal light ($\sim$64\%).  Table~\ref{tab:flux} reports the individual
flux contributions to the  total K-band emission (F$_{\rm{tot}}$). The
index 'dir' or 'sc' indicates whether it is direct or scattered light,
while  the index  'st' or  'th' designate  stellar and  thermal light,
respectively.  Assuming 
that the disk is made of silicate grains only, the fit to the AMBER data
  constrains the  grain size and consequently the  amount of scattered
  light. This  rules out solutions where only  direct thermal emission
  from the dust is assumed.  This means that scattered light can have
a major effect  in the overall  energy budget in  the NIR, and  a strong
impact on the inner disk  parameters inferred from modelling. This was
already  showed and taken  into account  in modelling  observations of
solar-mass young stars, as previously shown by \cite{pinte_4}, but
this is the first time that such case is made for a Herbig~Be star. 
This conclusion in fact contrasts with the usual assumption made to interpret
NIR interferometric  results in Herbig~AeBe  stars, \textit{i.e.} that
the  NIR  excess  emission  is  due only  to  direct  thermal
emission  from  hot  dust  located   at  the  inner  edge  of  a  disk
\citep{dullemond_1}.  In fact, 
scattering   or   additional   emitting   components  (such   as   gas
\citep{isella_1} or refractory  dust grains \citep{benisty_1}) seem to
be needed to reproduce  recent spectro-interferometric observations of
Herbig~AeBe stars.  Whether this  is relevant for  all the  objects of
this class should be validated on simpler cases.\\

\textbf{Evidence for sedimentation?} The  need for small grains at the
surface layers and  for large grains in the bulk of the disk
can be interpreted as resulting from grain growth and dust settling. A
toy model  was computed with dust  grain sizes
ranging from 0.05 to 3000~$\mu$m, a standard flaring
($\beta$=1.125),  but  considering a  larger  scale  height (36~AU  at
100~AU) and  a strong dust settling. The  predicted emission similarly
fits the SED.  Considering the age of the system, the disk is very likely
evolved and vertical stratification could well be happening. \\

\textbf{A gap and a tenuous inner disk:}
Modelling of the SED requires the presence of a gap.  In fact,
  a  continuous disk  model can  fit the  AMBER data  and the  flux at
  2.2~$\mu$m, but it produces  too much flux at 8.7~$\mu$m.  Similarly,
  a  continuous  disk model  that  would fit  the  MIDI  data and  the
  8.7~$\mu$m flux, would in turn predict too little flux in 
  the NIR.  This confirms the need for a gap, or at  least, for a strong
  discontinuity in  the radial density profile that  would produce the
  needed amounts of flux at specific locations.  

  If the need for this density discontinuity is clear,
  its size is nevertheless very model-dependent. In fact, we
  can not determine the position of the outer radius of the inner disk
  with precision, because it is fully resolved by AMBER.  The MIDI
observations constrain  the flux ratio  between the inner  and outer
  disks: when our inner disk extends up to $\sim$4~AU, the outer disk starts at
$\sim$13~AU. A change  in the inner disk geometry  would affect this
  estimate. In our model, it  is constrained by the temperature, hence
  position, of  the emitting particles. Considering  the typical grain
  size distributions given in Table~\ref{tab:modelparam}, our estimate
  is in good agreement with the values inferred from previous studies
\citep{brittain_1,grady_2}.

\citet{bouwman_1} required the disk to be more puffed up at
10~AU than  prediced by a standard  flared disk at the  same radius to
capture and re-emit enough light. They use 3.5~AU
as  scale height  for  dust at  $\sim$200~K  which is  much more  than
expected   for   gas    at   the   same   temperature   (\textit{i.e.}
1~AU).  \citet{brittain_1} found  that the  CO  disk extends  from
13$\pm$6~AU and  that at the inner edge  of the disk, the  gas is much
hotter ($\sim$1400~K) than the dust at the same radius. As the authors
suggest, this can be explained if there is a low density region in the
upper disk atmosphere, that results in a scale height of $\sim$3~AU. In
our model,  the scale height  of the outer  disk at its inner  edge is
$\sim$2.5~AU, due to  the small grains in the  surface layers, a value
in agreement with the predictions of \citet{brittain_1}.  

Moreover, \citet{bouwman_1} predicted a total dust mass of 10$^{25}$~g inside
the gap. This led \citet{brittain_1} to deduce from their estimate
of the  mass of  gas in the  inner disk ($\leq$3x10$^{25}$~g)  that the
gas-to-dust ratio is $\sim$3.  Our model requires a
total  dust    mass   of   3x10$^{-10}$~M$_{\odot}$   (6x10$^{23}$~g)
\textit{i.e.} a  gas-to-dust ratio of  $\sim$50. We therefore conclude that
these estimates  are highly  model-dependent and should  be considered
with care,  pending a  direct measurement of  the dust content  of the
inner disk.  The small accretion rate, inferred from UV spectroscopy
\citep{deleuil_1,grady_2} - a few times 10$^{-9}$M$_{\odot}$.yr$^{-1}$-
is enough to  empty the cavity in a few decades,  yet we are observing
dust in the inner disk at an age $\geq$10~Myr. Either we are observing
HD~100546 at a very special moment  in its evolution or the inner disk
is being replenished.  We note that a putative  large orbiting body, a
Jupiter-like planet located at about  10~AU, has the capacity to carve
the disk  and lead to such  a surface density  discontinuity at $\sim$13~AU,
while allowing  some material to  flow in continuously across  the gap
all the way to the star \citep{varniere_1}.

This transitional disk will be a primary target for ALMA that will
enable direct  determination of the gap extent,  by accurately probing
the gas surface density profile.  Before then, additional VLTI data
with three telescopes, allowing phase closure detections, will add new
constraints such as the asymmetry of the inner disk rim, the co-planarity
of the inner and  outer disks \citep{grady_3}, and whether
the star is off-center with respect to the gap \citep{grady_2}.

\begin{acknowledgements}
We  wish to  thank  ANR through  contract  ANR-07-BLAN-0221, the  PNPS
program   of   CNRS/INSU    (France),   and   INAF   (grant   ASI-INAF
I/016/07/0).   We   acknowledge   the   anonymous  referee   for   his
suggestions that improved the quality of the manuscript. 
\end{acknowledgements}

\bibliographystyle{aa}
\bibliography{hd100}

\end{document}